\documentclass[letter]{aa}
\usepackage{graphicx}
\usepackage{txfonts}
\usepackage{graphicx,epsfig}	
\usepackage{amsmath}           	
\usepackage{hyperref} 
\usepackage[T1]{fontenc}
\usepackage{ae,aecompl}
\usepackage{url}
\usepackage{booktabs}

\newcommand{\ppxf}{{\tt pPXF}\,}
\newcommand{\msun}{M$\mathrm{_\odot}$}

\newcommand{\Mstar}{$M_{*}$\,}

\newcommand{\re}{$R\rm_{e}$\,}

\begin{document} 
\title{Signs of `Everything Everywhere All At Once' formation in low surface brightness globular cluster-rich dwarf galaxies}
\author{A. Ferr\'e-Mateu\inst{1,2,3}, J. Gannon\inst{3,4}, D. Forbes\inst{3,4}, A.J. Romanowsky\inst{5,6}, M.L Buzzo\inst{3,4,7} \and J.P. Brodie\inst{3,4}}
\institute{Instituto de Astrof\'isica de Canarias, Av. Via Lactea s/n, E38205 La Laguna, Spain\\ 
\email{aferre@iac.es}
\and
Departamento de Astrof\'isica, Universidad de La Laguna, E-38200, La Laguna, Tenerife, Spain
\and 
Centre for Astrophysics \& Supercomputing, Swinburne University of Technology, Hawthorn VIC 3122, Australia
\and ARC Centre of Excellence for All Sky Astrophysics in 3 Dimensions (ASTRO 3D), Australia
\and 
Department of Physics \& Astronomy, San José State University, One Washington Square, San Jose, CA 95192, USA
\and 
Department of Astronomy \& Astrophysics, University of California Santa Cruz, 1156 High Street, Santa Cruz, CA 95064, USA
\and
European Southern Observatory, Karl-Schwarzschild-Strasse 2, 85748 Garching bei M\"unchen, Germany
\\
}
\vspace{-0.6cm}
\date{Received December 2024; accepted XXX, xxxx}
\abstract 
{Only two ultra-diffuse galaxies (UDGs) have spatially resolved stellar population properties, both showing radially flat-to-rising metallicity profiles, indicative of a different formation pathway to most dwarf galaxies. The scarcity of other low surface brightness (LSB) dwarfs with a similar analysis prevents a deeper understanding on this behaviour.}
{We investigate the radial profiles of the ages, metallicities and star formation histories of four globular cluster (GC) rich LSB dwarfs, newly observed within the \textit{Analysis of Galaxies At The Extremes} (AGATE) collaboration. DFX1 and DF07 are bona-fide UDGs, while PUDG-R27 and VCC~1448 are `nearly UDGs' (NUDGes). Comparing their and DF44's results to simulations, we aim to reveal their formation pathways.}
{We use \ppxf to fit different spectra extracted in annular apertures to recover the stellar population properties and compute their gradients. We compare those with a sample of literature classical dwarfs and simulations, in particular with simulated UDGs.}
{Our five LSB dwarfs present flat age and flat-to-rising metallicity profiles. The flat age gradients are compatible with those of classical dwarfs (both observed and from cosmological simulations), but the metallicity gradient diverges. All of our LSB dwarfs (except for PUDG-R27, showing a pronounced increasing metallicity profile) are compatible with being the extreme tail of the age--metallicity gradient relation, with a preference to co-eval formation forming the galaxy all at once.}
{This sample of GC-rich LSB dwarfs with spatially resolved properties confirms that they seem to follow a different formation path than classical dwarfs. However, larger samples with higher S/N spectra and varying amounts of GC richness are required to set robust constraints on the formation pathways of LSB dwarf galaxies.}

\keywords{galaxies: evolution -- galaxies: formation -- galaxies: kinematics and dynamics -- galaxies: stellar content }

\titlerunning{Stellar populations gradients of GC-rich LSB dwarfs}
\authorrunning{A.~Ferr\'e-Mateu et al.}
\maketitle
\section{Introduction}
Ultra-diffuse galaxies (UDGs) have provided a boost over the last decade in what we know about the low surface brightness (LSB) Universe and low-mass galaxies. UDGs have been described to have similar stellar masses to classical dwarfs (log(\Mstar/\msun)$\sim$7--8.5) but much larger effective radii (\re>1.5 kpc), and extremely low surface brightness ($\mu_{0,g}$>24 mag\,arcsec$^{-2}$; \citealt{vanDokkum2015a}). Forming these extreme dwarfs may be the result of several different scenarios. Some involve internal processes such as stellar feedback expansion, high spin or passive evolution (e.g. \citealt{DiCintio2017}; \citealt{Rong2017}), while others rely on external effects such as environmental quenching, galaxy mergers, ram-pressure or tidal stripping, and tidal heating (e.g. \citealt{Carleton2019}; \citealt{Tremmel2020}; \citealt{Fielder2024}). Moreover, combinations of both internal and external processes are highly probable (e.g.\ \citealt{Jiang2019}).

Such a variety of formation processes consequently lead to a variety of properties of the resulting UDG. Huge observational efforts have been carried out to obtain deep imaging to detect UDGs and derive robust sizes, morphologies and colours (e.g. \citealt{Venhola2017}; \citealt{Zaritsky2023}), or their GC populations (e.g. \citealt{Lim2020}; \citealt{Janssens2024}; \citealt{Marleau2024}). In addition, much telescope time has been devoted to the extremely challenging effort of obtaining spectra to derive stellar populations properties and internal kinematics to probe their dark matter halos (see \citet{Gannon2024b} for a compilation of spectroscopic data).

Overall, UDGs seem to follow the faint end of classical dwarfs in some aspects, while in others presenting intriguingly divergent properties, suggesting two distinct types (e.g.\ \citealt{Buzzo2024b}). The first type are thought to be classical dwarfs that puffed up due to one of the above scenarios. They should thus have stellar population properties resembling classical dwarfs: intermediate-to-old ages, relatively extended star formation histories, low metallicities that follow the standard mass--metallicity relation (MZR, e.g.\ \citealt{Ferre-Mateu2018a}; \citealt{Buzzo2022}; \hypertarget{FM+23}{\citealt{Ferre-Mateu2023}}, FM+23 hereafter) and a range of GC numbers. The second type of UDGs, dubbed `failed-galaxies', are those that do not follow the MZR of local dwarfs but instead that of high-redshift galaxies, with extremely low metallicities. These form very early on, have very old ages, and present the highest $\alpha$-abundance patterns (\hyperlink{FM+23}{FM+23}). They are expected to host, in general, the most populous GC systems \citep{Forbes2020}. 

\begin{figure*}
    \centering
    \includegraphics[width = 0.99 \textwidth]{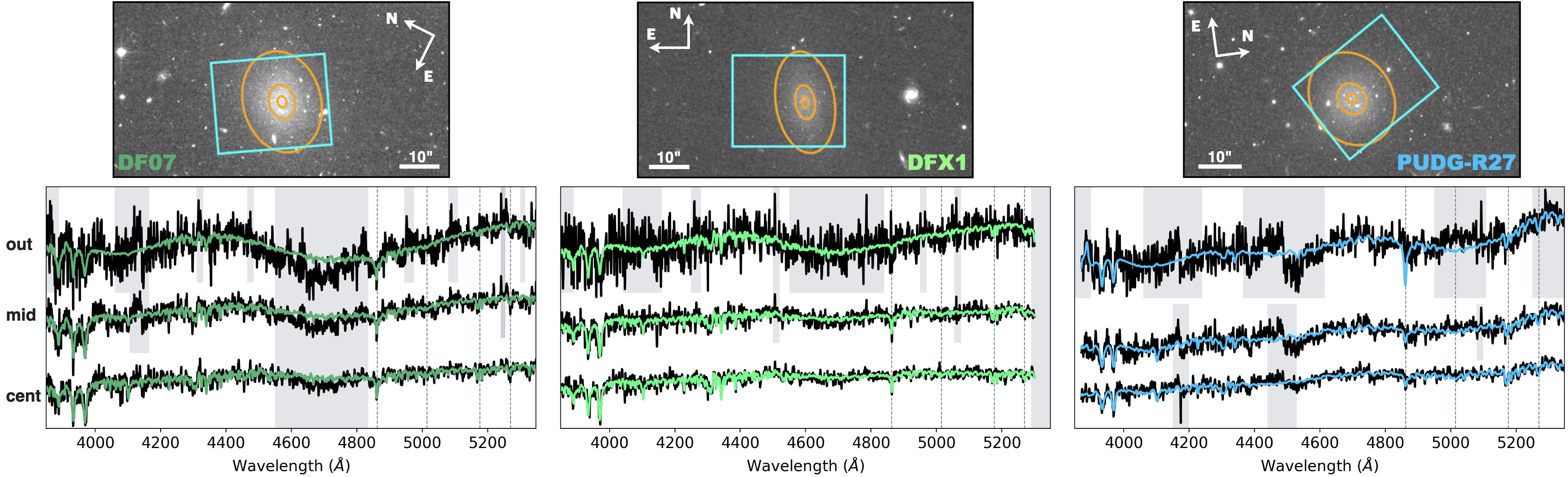}
    \vspace{-0.1cm}
    \caption{1$\times$1 arcmin cutouts around DF07, DFX1 and PUDG-R27 from \textit{Hubble Space Telescope} F814W data. The cyan rectangle indicates the positioning of the KCWI slicer, while the orange ellipses indicate the regions used for the extraction of spectra, corresponding to the three spectra below: central annulus (cent), middle aperture (mid), and outer annulus (out). For each spectrum, the coloured line corresponds to the \ppxf fit. Shaded regions show enforced masking, and dashed vertical lines mark some relevant lines: H$\beta$, Fe5015, Mg${_b}$ and Fe5270 (from left to right).}
    \end{figure*}
\label{fig:on_sky}

While simulations can reproduce some properties of individual observed UDGs, there is no simulation that can alone describe their observed large variety of properties. In \hyperlink{FM+23}{FM+23} we discussed the properties of UDGs from different cosmological simulations and  their caveats (e.g. too high metallicities in Illustris-TNG, \citealt{Nelson2018}; or not producing classical dwarfs in NIHAO, \citealt{DiCintio2017}). But one of the most striking discrepancies so far are their radial metallicity profiles. Simulations of isolated and cluster UDGs alike predict declining metallicity and flat age profiles towards the outer parts of the galaxies (e.g. \citealt{CardonaBarrero2023}; \citealt{Benavides2024}), similar to  classical dwarfs (e.g. \citealt{Mercado2021}). In contrast, the  only two observed UDGs to date with spatially resolved stellar populations (DF44, \citealt{vanDokkum2019b}; \citealt{Villaume2022}, and NGC1052-DF2, \citealt{Emsellem2019}; \citealt{Fensch2019}) have shown flat-to-rising profiles. 

Here we investigate the radial stellar populations trends of five cluster LSB dwarfs, including a re-analysis of DF44, within the \textit{Analysis of Galaxies At The Extremes} (AGATE) collaboration. Two of them are bona-fide UDGs according to the original definition of \citet{vanDokkum2015a} (DF07, DFX1). However, given the recent results of \citealt{Buzzo2024b}, showing that the distinct properties of the two classes of UDGs extend to other dwarf galaxies, we also include two Nearly UDGs (NUDGes, \citealt{Forbes2024}): PUDG-R27 (previously considered a UDG in \hyperlink{FM+23}{FM+23}), and VCC~1448 \citep{Gannon2024a}. All five LSB dwarfs are GC-rich (i.e. with more than 20 GCs), and we assume a $\Lambda$CDM cosmology with $H_0$=70 km s$^{-1}$ Mpc$^{-1}$, $\Omega_m$=0.27 and $\Omega_\Lambda$=0.73. 

\section{Data \& Methods}
\label{sect:data}
Observations and data reduction for DF07, DFX1 and PUDG-R27 are described in \hyperlink{FM+23}{FM+23}, which presented their global stellar population values. The data were obtained with the integral-field Keck Cosmic Web Imager (KCWI), and here we extract spectra from three annular apertures with semi-major axis radii: 0--1, 1--3 and 3--9\,arcsec, following each galaxy's photometric centre, position angle, and axis ratio (see Figure \ref{fig:on_sky}). Sky subtraction is performed using a 3$^{\prime\prime}$ annulus surrounding the extracted region. Figure~\ref{fig:on_sky} also shows the resulting spectra for each aperture (black line). The KCWI data for DF44 (re-analysed here) and VCC~1448 have higher signal-to-noise, allowing for more annuli (7 and 11, respectively). Details about these observations, data reduction and resulting spectra can be found in \citet{vanDokkum2019b} and \citet{Villaume2022} for DF44, and \citet{Gannon2024a} for VCC~1448. 

We follow established methods for deriving stellar populations from moderate S/N spectra of low-mass galaxies. In brief, we use \ppxf \citep{Cappellari2017)} to derive mean ages, metallicities and star formation histories (SFHs). We employ the E-MILES \citep{Vazdekis2015} single stellar population models with BaSTI isochrones and a \cite{kroupa2001} initial mass function, spanning 0.03 to 14 Gyr in age, and metallicities of [M/H] = $-$2.42 to $+$0.26 dex. The results from applying a regularization parameter or not yielded similar SFHs and stellar populations properties in the UDGs analysed in \hyperlink{FM+23}{FM+23}. We thus use a non-regularized solution for simplicity, with associated uncertainties obtained through Monte Carlo simulations of the fitting procedure (e.g. different configurations, varying regularization values, fitting polynomials). This procedure is performed for all five LSB galaxies, including DF44, for which we obtain results compatible with the published values of \citet{Villaume2022}.

\begin{figure*}
\centering
\includegraphics[width = 1.00 \textwidth]{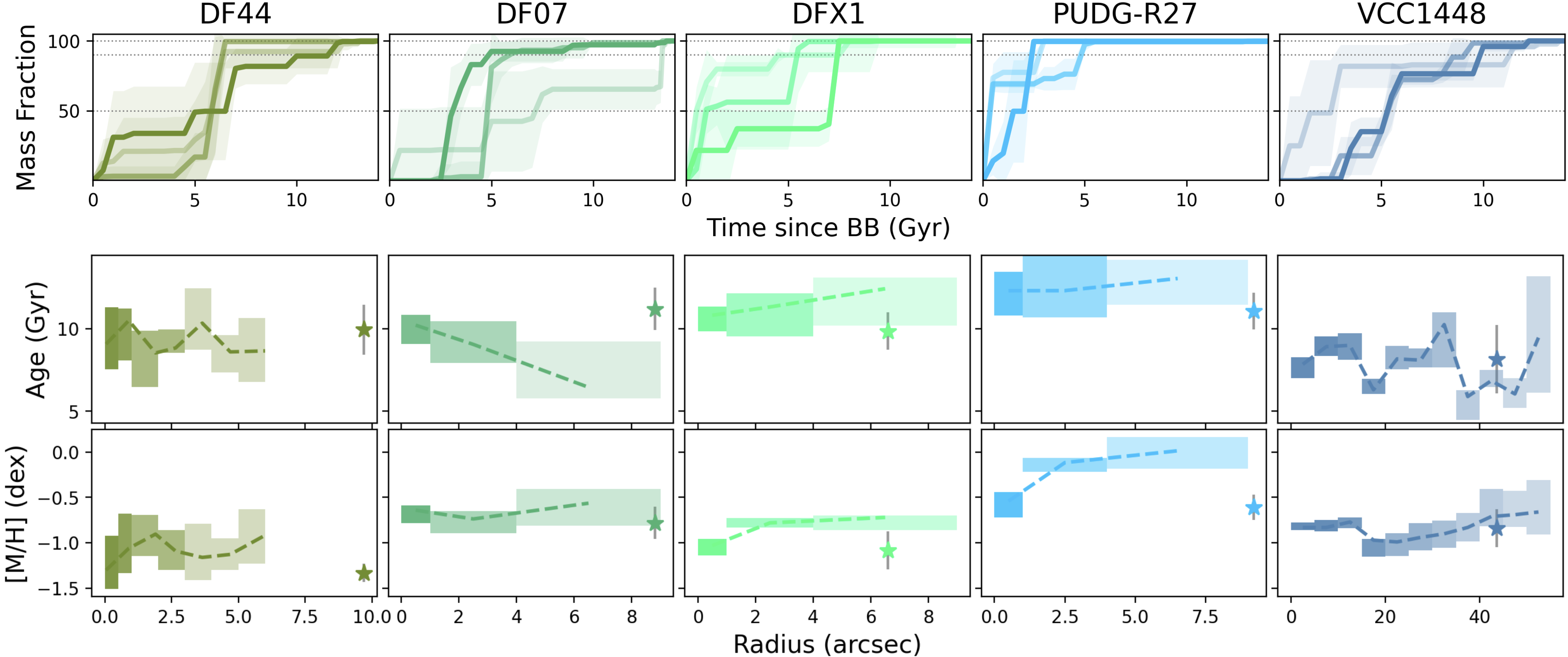}
\caption{\textit{Top}: SFHs of our LSB dwarfs, shown as the cumulative mass fraction assembled over cosmic time since the Big Bang. Dotted horizontal lines mark 50, 90 and 100\% of mass build-up. Lighter colours represent the outermost apertures (first, middle and outer bins for DF44 and VCC~1448 only), shaded regions show the fitting uncertainties. \textit{Middle and Bottom:} Mean mass-weighted age and metallicity profiles, with the different apertures and uncertainties represented by shaded regions (lighter shades for lower S/N). In each panel, a star symbol is shown at 1\re for the integrated value. Overall, the age profiles are flat with relatively old ages, whereas the metallicity profiles all show a flat-to-rising tendency.}
\label{fig:SFHgrads_fig}
\end{figure*}

\section{Stellar population gradients in LSB dwarfs}
The SFH for each bin is expressed in Figure~\ref{fig:SFHgrads_fig} as the mass fraction assembled since the Big Bang. Horizontal dotted lines mark certain percentages of mass fraction assembled, used to compute critical timescales, e.g. the quenching time as when the galaxy built 90\% of its mass. Independently of the initial time of formation, most of our LSB dwarfs formed relatively quickly. For instance, the three UDGs in Coma took $\sim$5\,Gyr to form half of their stellar mass and quenched $\sim$6--8\,Gyr ago. PUDG-R27 had instead an extremely early and fast formation, with very early quenching, indicative of very early cluster infall. VCC~1448 shows the most extended SFH in the sample, with a slow and steady formation history that quenched only $\sim$2\,Gyr ago, suggesting late infall into the Virgo Cluster. All of these quenching times are compatible with their respective phase space positions within their environment (\hyperlink{FM+23}{FM+23}; \citealt{Gannon2024a}).

We also show in the middle and bottom rows of Figure \ref{fig:SFHgrads_fig} the age and metallicity gradients, derived from the previous SFHs. We perform linear fits to these profiles, computed within 0.7\re. We also compute them within 1\re for PUDG-R27 and VCC~1448, given their larger coverage, although only VCC~1448 shows a change in their metallicity gradient, from $-$0.123$\pm$0.105\,(dex/\re) to 0.075$\pm$0.101\,(dex/\re). We find overall rather flat age profiles, with an average of $\bigtriangledown$log(Age)$\sim -0.003\pm$0.071\, (Gyr/\re), except for DF07 which shows a moderate decrease towards the outskirts. For the metallicity, we find flat to slightly rising profiles with on average $\bigtriangledown$[M/H]$\sim$0.123$\pm$0.061\,(dex/\re), similar to the literature results for DF44 and NGC1052-DF2. This is surprising as both nearby and local classical dwarfs show typically declining metallicity profiles (e.g. \citealt{Chilingarian2009}; \citealt{Weisz2014b}; \citealt{Taibi2022}). Table 1 summarises the main properties of each galaxy, as well as the measured gradients.

\begin{figure}
\centering
\includegraphics[scale=0.62]{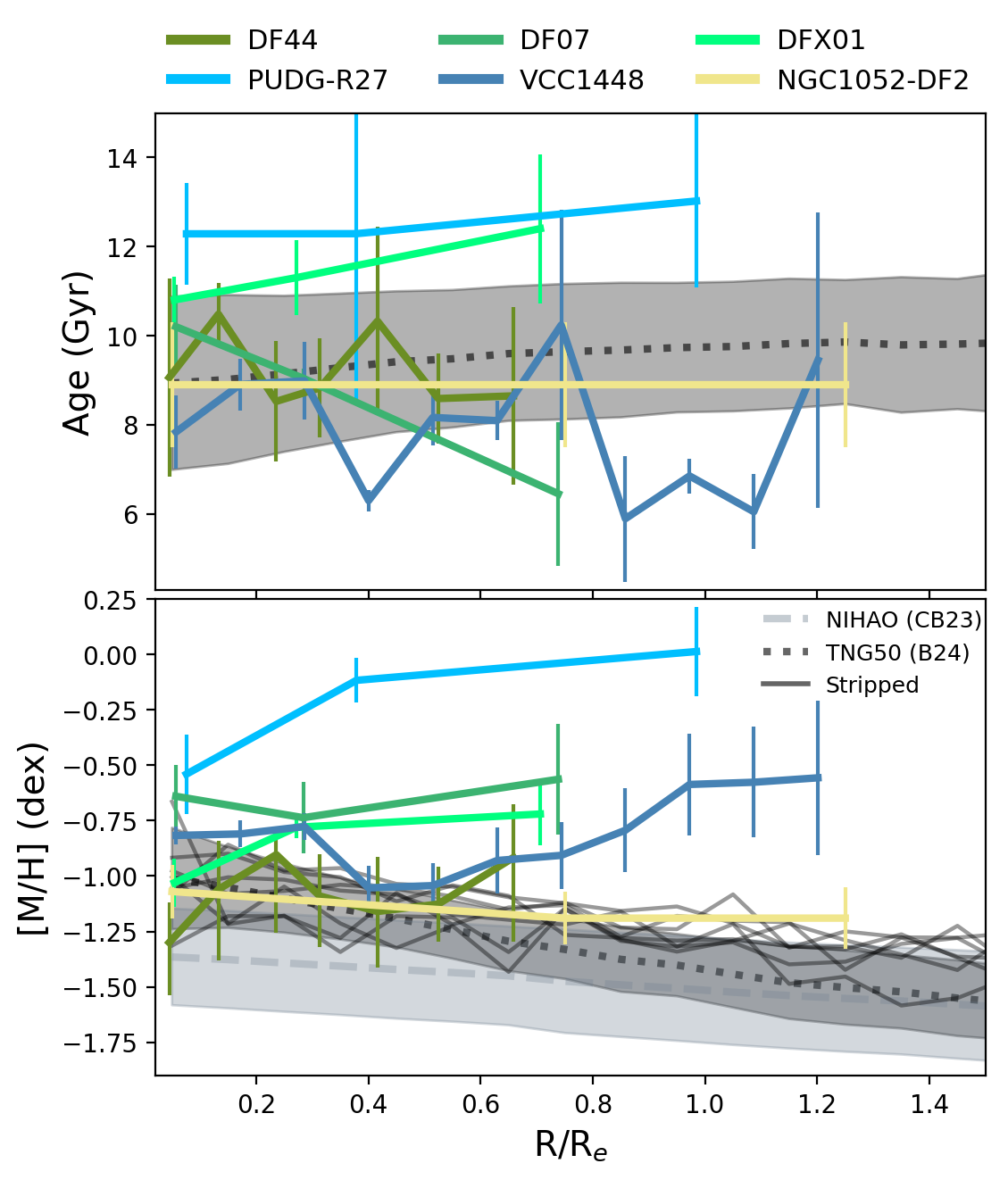}
\vspace{-0.35cm}
\caption{Age (top) and metallicity (bottom) profiles of observed LSB dwarfs (coloured lines), compared to the average trends of quiescent cluster UDGs in the TNG50 simulations (dotted lines and dark grey shading) and field UDGs from NIHAO (dashed line and light grey shade). Solid lines show tidally stripped UDGs from TNG50. While observed age profiles are mostly flat as in simulations, the metallicity profiles show a divergence, particularly at larger radii.}
\label{fig:grads_simul_fig}
\end{figure}

\begin{table*}
\centering
\label{table:grads}                      
\begin{tabular}{c |c c c c c c}   
\textbf{Galaxy}  & \textbf{\Mstar}   &  \textbf{\re} & \textbf{Age} & \textbf{[M/H]} & \textbf{$\bigtriangledown$log(Age)} & \textbf{$\bigtriangledown$log([M/H])}\\  
           &  ($\times 10^{8}$\msun) &   (kpc)       &    (Gyr)     &  (dex)         &      (Gyr/Re)                       &   (dex/Re)                           \\
\hline \hline
DF44     &  3.00 & 4.7 & 10.23$\pm$1.50 & $-$1.33$\pm$0.05 &  $-$0.003$\pm$0.031 &  0.123$\pm$0.082 \\     
DF07     &  4.35 & 4.3 & 11.18$\pm$1.27 & $-$0.78$\pm$0.18 &  $-$0.167$\pm$0.051 &  0.047$\pm$0.146 \\
DFX1     &  3.40 & 3.5 &  8.84$\pm$1.13 & $-$1.01$\pm$0.21 &     0.051$\pm$0.035 &  0.191$\pm$0.088 \\
PUDG-R27 &  4.84 & 2.1 & 11.04$\pm$0.61 & $-$0.61$\pm$0.14 &     0.022$\pm$0.060 &  0.508$\pm$0.077 \\
VCC1448  & 26.00 & 3.5 &  8.16$\pm$2.01 & $-$0.84$\pm$0.21 &  $-$0.020$\pm$0.205 & -0.160$\pm$0.123 \\ 
\midrule                                                   
\end{tabular}
\vspace{0.1cm}
\caption{\textit{Main properties of our LSB dwarfs:} Stellar masses (column 1), sizes (column 2), mean mass-weighted ages and metallicities (columns 3 and 4) from the literature (see main text); age (column 5) and metallicity (column 6) gradients computed within 0.7\re.}
\end{table*}

To better understand the age and metallicity gradients of the GC-rich LSB dwarfs in this study, we compare them in Figure \ref{fig:grads_simul_fig} to profiles from quenched cluster UDGs from the TNG50 simulation \citep{Benavides2024} and field UDGs from NIHAO \citep{CardonaBarrero2023}. These are shown as the mean (dotted and dashed lines, respectively) and standard deviation (shaded regions) for each simulation. These simulations have been previously shown to cover similar masses and sizes to observed UDGs (e.g. \citealt{Ferre-Mateu2018a}; \citealt{Benavides2024}). For completeness, we also include the profiles of the low-density environment UDG NGC1052-DF2 \citep{Fensch2019}, despite not being considered GC-rich. The top panel of Figure \ref{fig:grads_simul_fig} shows that the observed flat age profiles are overall very similar to those of UDGs in TNG50 (no age profiles were provided by NIHAO). 

Here we do not attempt to draw direct comparisons with the \textit{absolute} values of the metallicities (bottom panel), but rather with \textit{relative trends}, given the caveats intrinsic to each simulation. TNG50 produces systematically higher metallicities than observations (e.g.\ \citealt{Nelson2018}), which is usually solved by matching their metallicities to a given observed MZR at log(\Mstar/\msun)=9. This shift is $\sim-$0.4\,dex (\citealt{Sales2020}; \citealt{Benavides2024}), which we apply to our profiles here,noting that a different shift would not change the comparison. While no re-scaling is necessary for field NIHAO UDGs, these simulations produce much lower metallicities than most observed UDGs, only matched by UDGs in low-density environments (see \hyperlink{FM+23}{FM+23}). 

We find that, except for PUDG-R27 which shows a clear rising metallicity profile towards the outskirts, the rest of LSB dwarfs present flat profiles, at least up to $\sim$0.7\re. Such behaviour is similar to the gradients of NIHAO UDGs within 1\re\, which only start to decrease strongly farther out. Conversely, TNG50 UDGs present a steady strong decline already in the innermost regions (see \citealt{Benavides2024}). However, their central values are higher than NIHAO but similar to our observed ones. This match in central metallicities is most probably the result of studying similar environments (cluster). Interestingly, the observed flat trends are also found for tidally stripped UDGs in TNG50 (solid grey lines). Tidally stripped galaxies are expected to lie above the local MZR, while these LSB dwarfs mostly follow it (see \hyperlink{FM+23}{FM+23}), and thus a tidal origin is unlikely for all of them. We thus conclude that while the inner-most regions of our observed LSB dwarfs can show similar metallicity gradients to those in simulations, some cases are almost incompatible with being flat, let alone decreasing, in particular when reaching further out from the centre. Although the scatter in the observed dwarfs seems to be larger than in simulations, a similar dispersion is found if plotting all the individual gradients for simulated UDGs.

\section{Do LSB dwarfs form like classical dwarfs?}
\begin{figure*}
\centering
\includegraphics[scale=0.8]{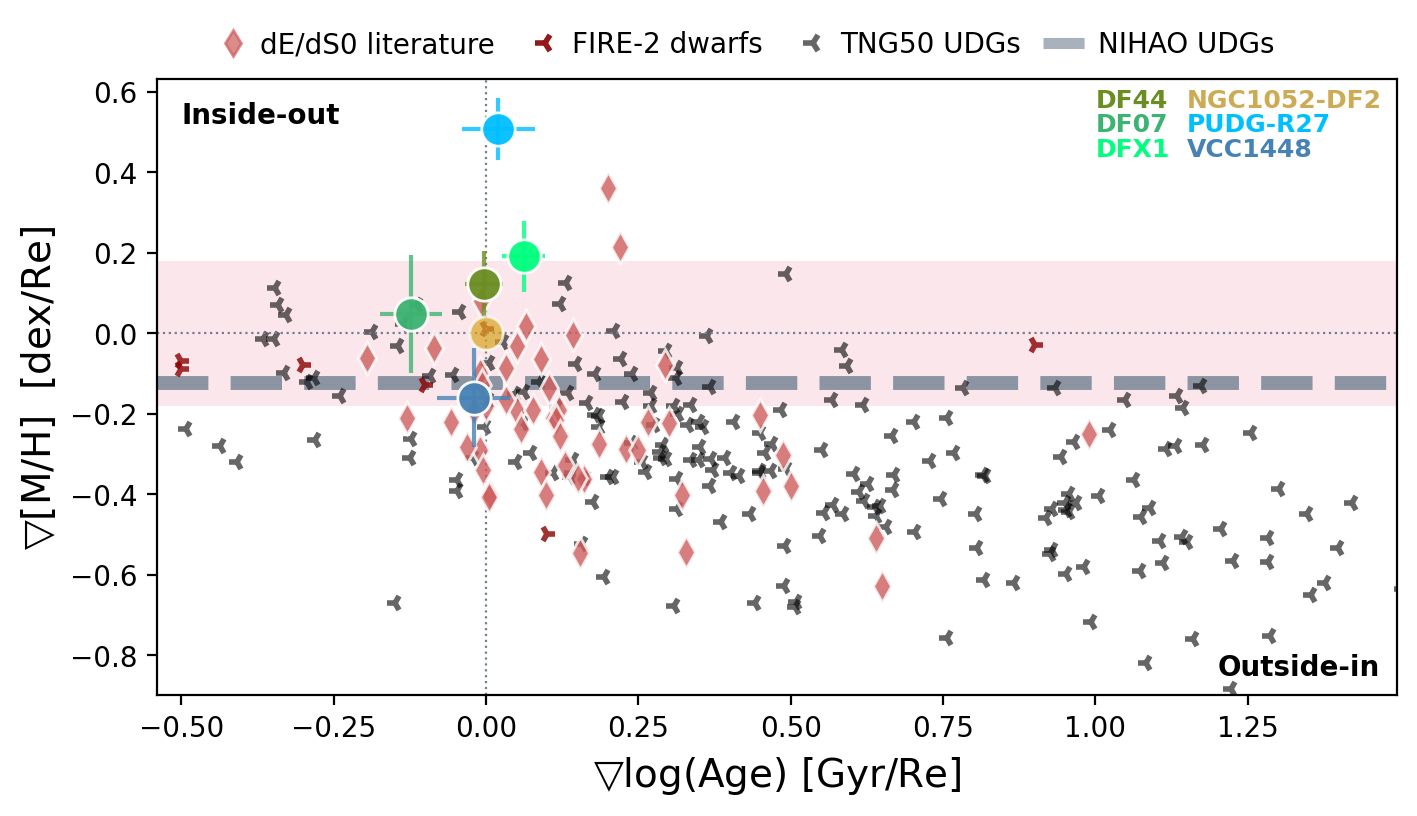}
\vspace{-0.25cm}
\caption{Age vs.\ metallicity gradients of the GC-rich LSB dwarfs in this study (coloured circles as in previous figures), now also including NGC1052-DF2. Quiescent simulated cluster UDGs from TNG50 (black 3-point asterisks), and the mean gradient (dashed grey) for field UDGs in NIHAO, both computed within 0.7\re\ to match our observations, are shown. Red 3-point asterisks correspond to simulated field dwarfs from the FIRE-2 simulations, marking in red shaded region the interval considered as `flat' according to \citet{Mercado2021}. Pink diamonds show observed classical dwarfs from the literature (see main text). Our LSB dwarfs are compatible with being the tail of the distribution of both observed and simulated dwarfs, around the intersection of zero-zero gradients. VCC~1448 shows the most similar behaviour to classical dwarfs if measured within 0.7\re, but we note that it would move closer to the rest of LSB dwarfs if computed further out. PUDG-R27 shows a strikingly different behaviour to all, hard to reconcile with any other existing dwarf galaxy. }
\label{fig:grads_agemet_fig}
\end{figure*}

Figure \ref{fig:grads_agemet_fig} shows the stellar population gradients of the cluster GC-rich LSB dwarfs from this work compared to a compilation of observed and simulated dwarfs of different types. We include dEs/dS0s, dSphs and tidal dwarfs from \citet{Koleva2011}, and Virgo dEs from \citet{Sybilska2017,Bidaran2022, Lipka2024a}. For the simulations we show field dwarfs from FIRE-2 \citep{Graus2019,Mercado2021}, marking with a shaded red region what is considered a `flat' metallicity gradient within those simulations. We also include simulated quenched cluster UDGs from TNG50 and the average metallicity gradient of UDGs from NIHAO (grey dashed line). For both simulations we have re-calculated their values within 0.7\re, to match our observations. 

Clearly, the bulk of classical dwarfs present positive age and negative metallicity gradients, favouring an outside-in formation. In this scenario, the star formation becomes more centrally concentrated with time (e.g.\ \citealt{Benitez-Llambay2016}), thus galaxies with early and fast SFHs should have flatter stellar population gradients than galaxies with ongoing SFHs (regardless of those being steady or bursty). Other simulations that have studied the origin of age gradients in dwarfs (e.g.\ \citealt{Riggs2024}) also support such outside-in formation, although they also found that some dwarfs had rather flat age profiles and in some cases, even decreasing. 

In contrast, Figure \ref{fig:grads_agemet_fig} shows that the LSB dwarfs in this work are populating the extreme tail of the distribution of dwarfs. Except for PUDG-R27, the majority of our LSB dwarfs reside around the zero-zero gradient intersection and within the shaded region of flat gradients. Together with the SFHs presented in Section \ref{sect:data}, this is indicative of an \textit{`everything everywhere all at once'}, co-eval formation for our observed LSB dwarfs. We note that VCC~1448 presents gradients more similar to classical dwarfs because its metallicity gradient within 0.7\re is slightly negative but becomes positive at larger radii, moving VCC~1448 closer to the rest of our LSB dwarfs. Noteworthy, a few observed and simulated classical dwarfs and UDGs share the same space in Figure \ref{fig:grads_agemet_fig} (e.g. FCC~288; \citealt{Koleva2011}). In fact, although scarce, some observed dwarfs have been found to have rising gas metallicity profiles (e.g.\ \citealt{Wang2019} and \citealt{Fu2024} for high-redshift dwarfs; or \citealt{Grossi2020} for two Virgo dwarfs). While this work has focussed in cluster LSB dwarfs, our results are similar to what has been found for featureless dwarfs in low-density regimes, which seem to be the equivalent to the NUDGes given their slightly higher brightness (e.g. \citealt{Lazar2024a,Lazar2024b}), which would indicate that this co-eval formation is not environment dependent, but we will require low-density LSB dwarfs to further investigate this.

\section{Conclusions}
We have studied the age and metallicity gradients of three GC-rich UDGs in clusters (DF44, DF07 and DFX1) and two NUDGes (nearly UDGs, PUDG-R27 and VCC~1448). We used \ppxf to fit apertures for each galaxy and obtain their SFHs, mean ages and metallicities at each radial bin. In general, we find that all five LSB dwarfs have formed relatively quickly (regardless of the time when they started forming stars), compatible with the location within their respective clusters, and in a co-eval way. PUDG-R27 shows the fastest and earliest quenching, while VCC~1448 presents a more extended SFH. They all present virtually flat age profiles with relatively old stellar populations ($\sim$9--10~Gyr), compatible with many observed and simulated LSB galaxies.

We perform a \textit{qualitative} comparison to the profiles of simulated UDGs, given that the latter present intrinsic caveats. While we find virtually flat age gradients, similar to simulations, all our LSB dwarfs show flat to rising metallicity profiles, in stark contrast to the simulated UDGs \citep{CardonaBarrero2020,Benavides2024}. We thus confirm the metallicity trends, first reported for DF44 in \citet{Villaume2022}, now with a larger sample of LSB dwarfs. 

Lastly, we have compared our stellar populations gradients to those of observed and simulated dwarfs in the literature, to see if they could share a common formation pathway. Our LSB dwarfs seem to be the extreme tail in the age--metallicity gradients relation of the classical dwarfs (both observed and simulated). Our LSB dwarfs tend to reside around the zero--zero gradient intersection and within the shaded region that indicates relatively flat gradients. We thus conclude that the most plausible formation path for GC-rich cluster LSB dwarfs is an \textit{`everything everywhere all at once'} co-eval formation, with some inside-out formation for the most extreme cases. Larger samples of LSB dwarfs covering different environments and GC populations, and with better data quality and larger spatial coverage, will thus be necessary to truly understand the formation pathways of these extreme dwarfs.

\begin{acknowledgements}
We wish to thank J.~Benavides and S.~Cardona-Barrero for providing the gradients in our required format, and A.~Villaume for the DF44 data. AFM has received support from RYC2021-031099-I and PID2021-123313NA-I00 of MICIN/AEI/10.13039/501100011033/FEDER,UE, NextGenerationEU/PRT. Part of this work was completed while JSG was an ECR visiting fellow at the IAC, he is grateful for their support. DF thanks the ARC DP20200102574, AJR the NSF grant AST-2308390, and we also acknowledge support from ASTRO-3D CE170100013.
This work was supported by a NASA Keck PI Data Award, administered by the NASA Exoplanet Science Institute. 
Data presented herein were obtained at the W. M. Keck Observatory from telescope time allocated to NASA through a scientific partnership with CalTech and the University of California. The authors wish to recognize and acknowledge the very significant cultural role that the summit of Maunakea has within the Indigenous Hawaiian community. We are most fortunate to have the opportunity to conduct observations from there.
\end{acknowledgements}

\bibliographystyle{aa}
\bibliography{UDGs_ref}
\end{document}